\documentclass[hyper]{JHEP} % 10pt is ignored!
\usepackage{graphicx,amssymb,bm,latexsym,amsmath}
\newcommand\fverb{\setbox\pippobox=\hbox\bgroup\verb}
\newcommand\fverbdo{\egroup\medskip\noindent%
            \fbox{\unhbox\pippobox}\ }

\newcommand\fverbit{\egroup\item[\fbox{\unhbox\pippobox}]}
\newbox\pippobox
\title{Spiky strings on $AdS_3\times S^3$ with NS-NS flux}
\author{Aritra Banerjee\\
Department of Physics, Indian Institute of Technology Kharagpur,\\
Kharagpur-721 302, India \\
Email: \email{aritra@phy.iitkgp.ernet.in}}
\author{Kamal L. Panigrahi\\
Department of Physics, Indian Institute of Technology Kharagpur,\\
Kharagpur-721 302, India , and \\
Department of Physics, CERN Theory Division, CH-1211, Geneva 23,
Switzerland\\
Email: \email{panigrahi@phy.iitkgp.ernet.in}}
\author{Pabitra M. Pradhan\\
Department of Physics, Indian Institute of Technology Kharagpur,\\
Kharagpur-721 302, India \\
Email: \email{ppabitra@phy.iitkgp.ernet.in}} \vskip .2in
\abstract{We study rigidly rotating strings in the background of
$AdS_3 \times S^3$ with Neveu-Schwarz (NS) fluxes. We find two
interesting limiting cases corresponding to the known giant magnon
and the new single spike solution of strings in the above
background and write down the dispersion relations among various
conserved charges. We use proper regularization to find the
correct relations among them. We further study the circular
strings and infinite spikes on AdS and study their properties.}

\begin{document}

\section{Introduction} AdS/CFT duality
\cite{Maldacena:1997re} \cite{Gubser:1998bc} \cite{Witten:1998qj}
relates spectrum of free strings in the string theory side to the
operator dimensions in the gauge theory in planar limit. Though
proving the duality has been a quite challenging subject primarily
due to that fact there are infinite tower of states in the string
theory side, in the last few years the conjectured duality has
been tested in various limits on both sides of the correspondence,
such as the large charge sectors
\cite{Berenstein:2002jq}\cite{Gubser:2002tv}\cite{Frolov:2003qc}\cite{Frolov:2003xy}.
With the realization that the idea of counting the operators from
gauge theory side can be elegantly formulated in terms of an
integrable spin chain \cite{Minahan:2002ve} , it has further been
established that integrability played an important role on both
sides of the duality, since the dual string theory also needs to
be integrable in the semiclassical limit \cite{Pohlmeyer:1975nb}
\cite{Minahan:2002rc} \cite{Tseytlin:2004xa} \cite{Hayashi:2007bq}
\cite{Okamura:2008jm}. In this connection, Hofman and
Maldacena(HM) put forward a special limit \cite{Hofman:2006xt},
using which determining the spectrum on both sides of the duality
became much easier. With these ideas, various large classes of
rigidly rotating and pulsating string solutions have been studied
in AdS and non-AdS backgrounds. These studies have also been
successful in providing a realization of the string states which
would correspond to some exact operators in the gauge theory side
(see, for example \cite{Lunin:2005jy} \cite{Bobev:2005cz}
\cite{Dorey:2006dq} \cite{Chen:2006gea} \cite{Arutyunov:2006gs}
\cite{Astolfi:2007uz} \cite{Chu:2006ae} \cite{Bobev:2006fg}
\cite{Kluson:2007qu} \cite{Lee:2008sk} \cite{David:2008yk}
\cite{Grignani:2008is} \cite{Lee:2008ui} \cite{Ryang:2008rc}
\cite{Benvenuti:2008bd} \cite{Abbott:2008qd}). In HM construction,
the spectrum on the field theory side consists of an elementary
excitation, the so called magnon which carries a momentum $p$
along the  finitely or infinitely long spin chain
\cite{Tseytlin:2003ii} \cite{Ishizeki:2007we}. The dual string
state, derived from the rigidly rotating string in the
$\mathbb{R}\times S^3$ presenting the same dispersion relation in
the large `t Hooft limit, is known as the giant magnon. However,a
more general class of rotating string solutions were also found
out in \cite{Kruczenski:2004wg} which are dual to a higher twist
operators in the boundary field theory. These kind of solutions
are called spiky strings. In addition to the rotating strings, the
spinning and folded strings have also been found out to have exact
correspondence with the dual operators in the gauge
theory\cite{Engquist:2003rn} \cite{Smedback:1998yn}. Indeed a
large class of rotating and pulsating string solution in various
backgrounds have been studied and the dual operators have also
been examined carefully. More recently string theory in the
background of $ AdS_3 \times S^3 \times T^4$ with mixed R-R and
NS-NS flux have been shown to be integrable and a S-matrix has
also been proposed \cite{Hoare:2013pma} \cite{Hoare:2013ida}
\cite{Hoare:2013lja}. Furthermore the giant magnon solution has
been studied solving the principal chiral model and the modified
dispersion relation has been presented to be \begin{eqnarray}
E-J_1 = \sqrt{(J_2 - qTp)^2 + 4 T^2 (1-q^2){\rm sin}^2
\frac{p}{2}} \ ,\label{I1}\end{eqnarray} where $p$ is the
worldsheet momentum and $T$ is the tension of the string and $q$
parametrizes the Neveu-Schwarz flux. It has been clear that for the
interpolating case of $q \neq 1$ the
spectrum cannot be detailed using either the WZW model approach
($q = 1$, pure NS-NS) or the Bethe ansatz approach ($q = 0$, pure R-R). Thus the
investigation of string states in the mixed flux backgrounds
was initiated expecting to find some relation between these two
approaches. However, this giant magnon solution (\ref{I1})
was expected to correspond to the appropriate operator in the dual
theory. The finite size corrections for this giant magnon has also
been proposed in \cite{Ahn:2014tua}. Further in
\cite{David:2014qta} more folded spinning string solutions in this
background have been studied. It was also shown, by using SO (2,2)
transformation and re-parametrization, that these spinning and
folded strings can be related to light like Wilson loops in
$AdS_3$ with Neveu-Schwarz flux. In this work we would like to
generalize these ideas further. First of all we solve the string
equations of motion in the background of $\mathbb{R} \times S^3$
with NS-NS flux and find two limiting case solutions, one
corresponding to the giant magnon \cite{Hoare:2013lja} and the
other corresponding to a new single spike solution. For these cases, the
dispersion relation among the Noether charges and deficit angle
gets modified  by a `shift term' proportional to the worldsheet
momenta due to the presence of the NS-NS flux, as shown in
\cite{Hoare:2013lja}. Further we generalize the magnon and spiky
string solutions by turning on NS-NS fields both in the $AdS$ and
sphere part, and study more general class of rotating open string solutions. We also
study several circular and infinite spike string configurations
and find the energy-spin relation with a discussion about the interplay
between them.

The rest of the paper is organized as follows. In
section-2 we study the general solution for the rigidly rotating
strings in the background $\mathbb{R}\times S^3$ with NS-NS flux
and find both giant magnon and single spike solutions as two
limiting cases. Section-3  is devoted to the study of rotating,
circular and helical open strings in the background of $AdS_3 \times
S^3$ with NS-NS flux both on AdS and on sphere. Finally in
section-4 we present our conclusions.

\section{Spiky strings in $\mathbb{R} \times S^3$ with NS-NS flux}
We start with the metric and the background field for $AdS_3
\times S^3 \times T^4$ geometry supported by the NS-NS fluxes.
This geometry can be obtained by taking the near horizon limit of
the $NS1-NS5$ background, see for example
\cite{Papadopoulos:1999tw}. The full $AdS_3 \times S^3$ metric
with NS-NS B-fields is as follows
\begin{eqnarray}
ds^2 &=& -\cosh^2\rho dt^2 + d\rho^2 + \sinh^2\rho d\phi^2 +
d\theta^2 + \sin^2\theta d\phi_1^2 + \cos^2\theta d\phi_2^2, \cr &
\cr &&
b_{t\phi}=q\sinh^2\rho,~~~~~~~~~~b_{\phi_1\phi_2}=-q\cos^2\theta.
\label{1}
\end{eqnarray}
In this section we wish to review the giant magnon solution
proposed in \cite{Hoare:2013lja} and find a new single spike
solution of string equations of motion in the background of
$\mathbb{R} \times S^3$ with NS-NS B-field. The relevant metric
and background field is given by (putting $\rho = 0$)
\begin{equation}
ds^2 = - dt^2 + d\theta^2 + \sin^2\theta d\phi_1^2 + \cos^2\theta
d\phi_2^2, ~~~~~~~~~~b_{\phi_1\phi_2}=-q\cos^2\theta. \label{A1}
\end{equation}
The Polyakov action of a fundamental string in the conformal gauge
for the said background can be written as
\begin{eqnarray}
S = \frac{T}{2}\int d\tau d\sigma\Big[-(\dot{t}^2 -
{t^{\prime}}^2) &+& \dot{\theta}^2  - {\theta^{\prime}}^2 +
\sin^2\theta(\dot{\phi_1}^2 - {\phi_1^{\prime}}^2) \cr & \cr && +
\cos^2\theta(\dot{\phi_2}^2 - {\phi_2^{\prime}}^2)  -
2q\cos^2\theta (\dot{\phi_1}{\phi_2}^{\prime} - {\phi_1}^{\prime}
\dot{\phi_2})\Big], \label{A2}
\end{eqnarray}
where the `dot' and `prime' denote the derivatives with respect to
$\tau$ and $\sigma$ respectively and $T =
\frac{\sqrt{\lambda}}{2\pi}$, where $\lambda$ is the `t Hooft
coupling constant. We write the anstaz for the rigidly rotating
string as
\begin{eqnarray}
t = \kappa \tau ,~~~ \theta = \theta(y), ~~~  \phi_1 &=&
\omega_1[\tau + g_1(y)],  ~~~ \phi_2 = \omega_2[\tau + g_2(y)],
\label{A3}
\end{eqnarray}
where $y = a\sigma - b\tau$. The equations of motion for $\phi_1$
and $\phi_2$ give the following
\begin{eqnarray}
g_{1y} &=& \frac{1}{a^2 - b^2}\left[\frac{A_1+a\omega_2
q}{\omega_1 \sin^2\theta} -aq\frac{\omega_2}{\omega_1} - b
\right], \cr & \cr g_{2y} &=& \frac{1}{a^2 -
b^2}\left[\frac{A_2}{\omega_2\cos^2\theta}
-aq\frac{\omega_1}{\omega_2} - b \right], \label{A4}
\end{eqnarray}
where $g_y = \frac{\partial g}{\partial y}$. Solving the equation
of motion for $\theta$, we get
\begin{equation}
(a^2-b^2)^2\theta_y^2 =
-a^2(1-q^2)(\omega^2_1-\omega_2^2)\sin^2\theta - \frac{(A_1+a
\omega_2 q)^2}{\sin^2\theta} - \frac{A^2_2}{\cos^2\theta} + C,
 \label{A5}
\end{equation}
where $C$ is an integration constant. Invoking the boundary
condition $\frac{\partial \theta}{\partial y} \rightarrow 0$ at
$\theta = \frac{\pi}{2}$, we get
\begin{equation}
A_2 = 0, \>\>\> C = a^2(1-q^2)(\omega^2_1-\omega_2^2) + (A^2_1 + a
q \omega_2)^2. \label{A6}
\end{equation}
Putting the equation (\ref{A5}) in the Virasoro constraint
$T_{\tau\sigma}=0$ we get
\begin{equation}
C = \frac{\omega_1}{b}(a^2+b^2)(A_1+aq\omega_2) +
2aq\omega_2(A_1+aq\omega_2)-\left[\frac{1}{b}aq\omega_1\omega_2(a^2+b^2)+a^2(q^2\omega^2_1+\omega^2_2)\right].
\label{A7}
\end{equation}
Equating (\ref{A6}) and (\ref{A7}), we get a quadratic equation
for $A_1$ and solving that we get the following limiting values of
$A_1$
\begin{eqnarray}
A_1 &=& b\omega_1~~~~~~~~{\rm Giant \> Magnon\> solution}\\
&=& a^2\frac{\omega_1}{b}~~~~~~{\rm Single \> Spike\> solution}
\label{A8}
\end{eqnarray}
\subsection{First limiting case: Giant Magnon solution}
Using $A_1 = b\omega_1$ in the equation (\ref{A5}) and (\ref{A6}),
we get
\begin{equation}
\theta_y = \frac{\Omega}{a^2-b^2}\cot\theta \sqrt{\sin^2\theta -
\sin^2\theta_0} \ , \label{A9}
\end{equation}
where, ${\Omega}^2=a^2(1-q^2)(\omega^2_1-\omega^2_2)$ and
$\sin\theta_0 = \frac{b\omega_1 + aq\omega_2 }{\Omega}$. Now the
conserved charges and deficit angle become
\begin{eqnarray}
E &=& \kappa T\int d\sigma, \cr & \cr  J_1 &=&
\frac{T}{a^2-b^2}\int d\sigma~[\omega_1 a^2(1-q^2)\sin^2\theta +
a^2q^2\omega_1- \omega_1 b^2 ], \cr & \cr J_2 &=&
\frac{T}{a^2-b^2} \int d\sigma~ \left[\omega_2 a^2(1
-q^2)+(a^2q^2\omega_2+abq\omega_1)
\frac{1}{\sin^2\theta}\right]\cos^2\theta, \cr & \cr \Delta\phi_1
&=& \frac{b\omega_1+aq\omega_2}{a^2-b^2} \int dy
~~~\frac{\cos^2\theta}{\sin^2\theta}. \label{A10}
\end{eqnarray}
Here, we note that integrals corresponding to both
$E$ and $J_1$ diverge, however their difference is
finite. The above charges can be shown to satisfy the following relation
\begin{equation}
\frac{E}{\kappa} - \frac{J_1}{\omega_1} = \frac{J_2-qT
\Delta\phi_1}{\omega_2} . \label{A11}
\end{equation}
Evaluating the right side of the above equation we get
\begin{equation}
\frac{J_2-qT\Delta\phi_1}{\omega_2} =
\frac{1}{\omega_1}\sqrt{(J_2-qT\Delta\phi_1)^2 +
4T^2(1-q^2)(1-z_0^2)}, \label{A12}
\end{equation}
where $z_0=\sin\theta_0$ and $\Delta\phi_1=\pi-2\theta_0$. Now we
can rewrite the equation (\ref{A11}) as
\begin{eqnarray}
\tilde{E} - J_1 = \sqrt{(J_2-qT\Delta\phi_1)^2 +
4T^2(1-q^2)\sin^2{\frac{\Delta \phi_1}{2}}}, \label{A13}
\end{eqnarray}
where $\tilde{E} = \frac{\omega_1}{\kappa}E$. This dispersion
relation (\ref{A13}) was obtained in \cite{Hoare:2013lja} by
solving the principal chiral model along with a Wess-Zumino term
in this background. It is evident that the presence of the term
linear in $\Delta \phi_1$ (which is identified to the worldsheet momentum $p$)
spoils the periodicity of the dispersion
relation, which is a completely new feature of these solutions.
Although, the implication of this in the dual picture is not clear.

It is quite evident that the dispersion relation reduces to the
usual dyonic giant magnon relation \cite{Ishizeki:2007we} in the
$q= 0$ case, which is interpreted as a bound state of $J_2$ number
of elementary magnons. In the other limit $q=1$, it was shown in
\cite{Hoare:2013lja} that with $J_2 = 1$, the magnon dispersion
relation simply corresponds to the  energy state $\epsilon = 1 -
Tp$, which occurs in the perturbative S-matrix as discussed in
\cite{Hoare:2013ida}.
\subsection{Second limiting case: Single spike solution}
Now we take $A_1 = a^2\frac{\omega_1}{b}$, which makes the
$\theta$ equation
\begin{equation}
\theta_y = \frac{\Omega}{a^2-b^2}\cot\theta \sqrt{\sin^2\theta -
\sin^2\theta_1} \ , \label{A14}
\end{equation}
where, $\sin\theta_1 = \frac{a^2\omega_1 + abq\omega_2
}{b\Omega}$. The conserved charges and deficit angles are
\begin{eqnarray}
E &=& \kappa T\int d\sigma, \cr & \cr  J_1 &=&
-\frac{T}{a^2-b^2}\int d\sigma~\omega_1 a^2(1-q^2)\cos^2\theta ,
\cr & \cr J_2 &=& \frac{T}{a^2-b^2} \int d\sigma~ \left[\omega_2
a^2(1 -q^2)+(a^2q^2\omega_2+a^3q\frac{\omega_1}{b})
\frac{1}{\sin^2\theta}\right]\cos^2\theta, \cr & \cr \Delta\phi_1
&=& \frac{a}{a^2-b^2} \int dy
~~~\left[\frac{a^2\omega_1+abq\omega_2}{b\sin^2\theta}-(aq\omega_2+b\omega_1)\right].
\label{A15}
\end{eqnarray}
From above charges we can see that while both $E$ and $\Delta\phi_1$
diverge, the combination $E-T\Delta\phi_1$ remains finite, and can be
written as
\begin{equation}
E-T\Delta\phi_1=2T(\frac{\pi}{2}-\theta_1). \label{A16}
\end{equation}
The dispersion relation between the angular momenta can be
written as follows,
\begin{equation}
J_1= \sqrt{[J_2-qT(\Delta\phi_1)_{reg}]^2 +
4T^2(1-q^2)\sin^2\frac{(\Delta\phi_1)_{reg}}{2}}, \label{A17}
\end{equation}
where
\begin{eqnarray}
J_1 &=& 2T\omega_1~ \sqrt{\frac{1-q^2}{\omega^2_1-\omega^2_2}}
~\cos\theta_1, \cr & \cr J_2 &=& -2T\omega_2~
\sqrt{\frac{1-q^2}{\omega^2_1-\omega^2_2}} ~\cos\theta_1 - 2q T
(\pi-2\theta_1), \cr & \cr (\Delta\phi_1)_{reg} &=&
-2\cos^{-1}\sin\theta_1 , \label{A18}
\end{eqnarray}
And the expression of $\Delta\phi_1$ has been regularized by the energy in the
equation (\ref{A15}). These relations (\ref{A16}) and (\ref{A17})
can be thought of as the generalized spike solution in the
presence of background flux. Note that by putting $q =0$ one
arrives at the single spike solution obtained in
\cite{Ishizeki:2007we}. Of course the $q = 1$ limit of the above solution
remains to be understood in a better way as before.
\section{Strings in $AdS_3\times S^3$ with Neveu-Schwarz flux}
In this section we will generalize the solutions presented in the
last section by turning on one spin $S$ along AdS and two angular
momenta $J_1, J_2$ along the $S^3$ of the full geometry
$AdS_3\times S^3$ with Neveu-Schwarz fluxes turned on both in the
AdS and the sphere part as written in (\ref{1}). The Polyakov
action of the fundamental string in conformal gauge for this background is written as
\begin{eqnarray}
S &=& \frac{T}{2}\int d\tau d\sigma\Big[-\cosh^2\rho (\dot{t}^2 -
{t^{\prime}}^2) + \dot{\rho}^2 - {\rho^{\prime}}^2 +
\sinh^2\rho(\dot{\phi}^2 - {\phi^{\prime}}^2) + \dot{\theta}^2 \cr
& \cr && - {\theta^{\prime}}^2 + \sin^2\theta(\dot{\phi_1}^2 -
{\phi_1^{\prime}}^2) +  \cos^2\theta(\dot{\phi_2}^2 -
{\phi_2^{\prime}}^2)  +  2q\sinh^2\rho (\dot{t}\phi^{\prime} -
t^{\prime}\dot{\phi}) \cr & \cr && - 2q\cos^2\theta
(\dot{\phi_1}{\phi_2}^{\prime} - {\phi_1}^{\prime}
\dot{\phi_2})\Big]. \label{2}
\end{eqnarray}
We take the following anstaz to parameterize the three spin
rigidly rotating open strings which have one spin $S$ in the AdS and two
angular momenta in the $S^3$ part as,
\begin{eqnarray}
t &=& \tau + h_1(y), ~~~ \rho = \rho(y), ~~~ \phi = \omega_1[\tau
+ h_2(y)], \cr & \cr \phi_1 &=& \tau + h_3(y), ~~~ \theta =
\theta(y), ~~~ \phi_2 = \omega_2[\tau + h_4(y)]. \label{3}
\end{eqnarray}
Solving the equations of motion for $t, \phi, \phi_1$ and
$\phi_2$, we have the following differential equations for $h_1,
h_2, h_3$ and $h_4$
\begin{eqnarray}
h_{1y} &=& \frac{1}{a^2 - b^2}\left[\frac{A_1-a\omega_1
q}{\cosh^2\rho} + a\omega_1 q - b\right], \cr & \cr h_{2y} &=&
\frac{1}{a^2 - b^2}\left[\frac{A_2}{\sinh^2\rho} +
\frac{aq}{\omega_1} - b\right], \cr & \cr h_{3y} &=& \frac{1}{a^2
- b^2}\left[\frac{A_3+a\omega_2 q}{\sin^2\theta} -a\omega_2 q - b
\right], \cr & \cr h_{4y} &=& \frac{1}{a^2 -
b^2}\left[\frac{A_4}{\cos^2\theta} -\frac{aq}{\omega_2} - b
\right], \label{4}
\end{eqnarray}
where $h_y = \frac{\partial h}{\partial y}$. Also, The
substraction of Virasoro constraints $T_{\tau\sigma} = 0$ and
$T_{\tau\tau} + T_{\sigma\sigma}  = 0$ gives the following
relation among various constants appearing in the differential
equations
\begin{equation}
-A_1 +  A_2\omega^2_1 + A_3 +  A_4 \omega^2_2 = 0. \label{5}
\end{equation}
Now, the equation of motion for $\rho$ and $\theta$ become
\begin{equation}
(a^2-b^2)\rho_{yy} + \frac{1}{a^2-b^2} \sinh\rho\cosh\rho
\left[\frac{(A_1-a\omega_1 q)^2}{\cosh^4\rho}  - \frac{\omega^2_1
A_2^2} {\sinh^4\rho} - a^2(1-q^2)(1-\omega^2_1)\right] = 0,
\label{6}
\end{equation}
\begin{equation}
(a^2-b^2)\theta_{yy} + \frac{1}{a^2-b^2} \sin\theta\cos\theta
\left[a^2(1-q^2)(1-\omega^2_2) - \frac{(A_3+a\omega_2
q)^2}{\sin^4\theta}  + \frac{ \omega^2_2 A_4^2}
{\cos^4\theta}\right] = 0, \label{7}
\end{equation}
where $\rho_{yy} = \frac{\partial^2 \rho}{\partial y^2}$ and
$\theta_{yy} = \frac{\partial^2 \theta}{\partial y^2}$. We can then
write the conserved charges as
\begin{eqnarray}
E &=& T~\int d\sigma~\left[(1 - bh_{1y}) \cosh^2\rho-a \omega_1 q~
h_{2y} \sinh^2\rho\right], \cr & \cr S &=& T~\int d\sigma~\left[
\omega_1(1 - b h_{2y})\sinh^2\rho - aq ~h_{1y} \sinh^2\rho\right]
, \cr & \cr J_1 &=& T~\int d\sigma~\left[(1 - bh_{3y})\sin^2\theta
- a\omega_2 q~ h_{4y} \cos^2\theta\right], \cr & \cr J_2 &=& T
~\int d\sigma~\left[\omega_2(1 - b h_{4y})\cos^2\theta + aq~h_{3y}
\cos^2\theta \right]  . \label{8}
\end{eqnarray}
Next, by choosing particular value of the constants as in the previous section,
we would like to look for giant magnon and single spike
solutions for the string and further look for the circular and
helical string configurations.
\subsection{Giant Magnon Solution}
For finding out the giant magnon solution, we choose the
integration constants as  $A_1 = b = A_3$ and $A_2 = 0 = A_4$. The
solution of equations (\ref{6}) and (\ref{7}) become
\begin{equation}
\rho_y^2 = \frac{1}{(a^2-b^2)^2} \left[a^2(1-q^2)(1-\omega_1^2)
-\frac{(b-a\omega_1 q)^2}{\cosh^2\rho}\right]\sinh^2\rho,
\label{9}
\end{equation}
and
\begin{equation}
\theta_y^2 = \frac{1}{(a^2-b^2)^2}\left[a^2(1-q^2)(1-\omega_2^2) -
\frac{(b+a \omega_2 q)^2}{\sin^2\theta}\right] \cos^2\theta \ .
\label{10}
\end{equation}
Note that the above equations can be rewritten as
\begin{equation}
\rho_y = \frac{\Omega_1}{a^2-b^2}\tanh\rho \sqrt{\cosh^2\rho -
\cosh^2\rho_0} \ , \label{11}
\end{equation}
and
\begin{equation}
\theta_y = \frac{\Omega_2}{a^2-b^2}\cot\theta \sqrt{\sin^2\theta -
\sin^2\theta_0} \ , \label{12}
\end{equation}
where, ${\Omega_1}^2=a^2(1-q^2)(1-\omega^2_1),~~~\Omega^2_2 = a^2
(1-q^2)(1-\omega^2_2),~~~\cosh\rho_0 = \frac{b-a\omega_1
q}{\Omega_1}$, and  $\sin\theta_0 = \frac{b+a\omega_2
q}{\Omega_2}$. Now the conserved charges (\ref{8}) and deficit
angles become
\begin{eqnarray}
E &=& \frac{T}{a^2-b^2}\int d\sigma~[a^2(1-q^2)\cosh^2\rho
+a^2q^2- b^2], \cr & \cr \frac{S}{\omega_1} &=&
\frac{T}{a^2-b^2}\int d\sigma~ \left[a^2(1-q^2) +
(a^2q^2-\frac{abq}{\omega_1})\frac{1}{\cosh^2\rho}\right]\sinh^2\rho,
\cr & \cr J_1 &=& \frac{T}{a^2-b^2}\int
d\sigma~[a^2(1-q^2)\sin^2\theta + a^2q^2 - b^2], \cr & \cr
\frac{J_2}{\omega_2} &=& \frac{T}{a^2-b^2} \int d\sigma~
\left[a^2(1 -q^2)+(a^2q^2+\frac{abq}{\omega_2})
\frac{1}{\sin^2\theta}\right]\cos^2\theta, \cr & \cr \Delta\phi_1
&=& \frac{1}{a^2-b^2} \int dy ~~(b+a\omega_2
q)~\frac{\cos^2\theta}{\sin^2\theta}, \cr & \cr \Delta t &=&
\frac{1}{a^2-b^2} \int dy ~~(a\omega_1 q - b)~
\frac{\sinh^2\rho}{\cosh^2\rho}. \label{13}
\end{eqnarray}
Here $\Delta t$ is the time difference between two endpoints of the string.
It is clear from the above expressions that we have the following
relation among various conserved charges
\begin{equation}
E - J_1 = \frac{S-qT\Delta t}{\omega_1} + \frac{J_2-qT
\Delta\phi_1}{\omega_2}. \label{14}
\end{equation}
As the conserved charges are divergent, we regularize them to
remove the divergent part. Let us write
\begin{eqnarray}
\frac{S-qT\Delta t}{\omega_1} =
\frac{2Ta}{a^2-b^2}\int_0^{\infty}~ \frac{d\rho}{\rho_y} ~ (1-q^2)
\sinh^2\rho = 2T \sqrt{\frac{1-q^2}{1-\omega_1^2}}
\int_{1}^{\infty}~ dz \frac{z}{\sqrt{z^2 - z_0^2}} , \label{15}
\end{eqnarray}
where $z = \cosh\rho$ and $z_0 = \cosh\rho_0 = \frac{b-a\omega_1
q}{\Omega_1}$. From the equations (\ref{13}) and (\ref{14}), we
can see that $E$ and $J_1$ have divergences in the IR whereas
$E-J_1$ has a divergence in UV which corresponds to $\rho =
\infty$ i.e the AdS boundary. So the divergence can be canceled by
generating a counter term by deforming the integration contour
away from $\rho = \infty$. This can be visualized in the sense
that these strings never reach out to the boundary of $AdS_3$.
Introducing a hard cut-off to regulate the UV divergence in the
integral and following the prescription in \cite{Minahan:2006bd} ,
we have the regularized value given as
\begin{equation}
\frac{{(S-qT\Delta t)}_{reg}}{\omega_1} = -2T~
\sqrt{\frac{1-q^2}{1-\omega_1^2}}\sqrt{1-z_0^2} \ . \label{16}
\end{equation}
From the above expression, we find the following relation
\begin{equation}
\frac{{(S-qT\Delta t)}_{reg}}{\omega_1} = - \sqrt{{(S-qT\Delta
t)}^2_{reg} + 4T^2(1-q^2)(1-z_0^2)} \ . \label{17}
\end{equation}
Similarly it is evident that
\begin{equation}
\frac{J_2-qT \Delta\phi_1}{\omega_2}
=\frac{2Ta}{a^2-b^2}\int_{\theta_0}^{\frac{\pi}{2}}~
\frac{d\theta}{\theta_y} ~ (1-q^2) \cos^2\theta = 2T
\sqrt{\frac{1-q^2}{1-\omega_2^2}} \int_{x_0}^{1}~ dx
\frac{x}{\sqrt{x^2 - x_0^2}} \ , \label{18}
\end{equation}
where $x = \sin\theta$ and $x_0 = \sin\theta_0 = \frac{b+a\omega_2
q}{\Omega_2}$. Which gives
\begin{equation}
\frac{J_2-qT\Delta\phi_1}{\omega_2} = 2T~
\sqrt{\frac{1-q^2}{1-\omega_2^2}}\sqrt{1-x_0^2} \ . \label{19}
\end{equation}
Further, the above expression can be written as
\begin{equation}
\frac{J_2-qT\Delta\phi_1}{\omega_2} = \sqrt{(J_2-qT\Delta\phi_1)^2
+ 4T^2(1-q^2)(1-x_0^2)}. \label{20}
\end{equation}
Now the time difference is given by
\begin{eqnarray}
\Delta t &=& 2~ \frac{a\omega_1 q - b}{\Omega_1}~
\int_{0}^{\infty}~d\rho~\frac{\tanh\rho}{\sqrt{\cosh^2\rho -
\cosh^2\rho_0}} =  - 2 \tan^{-1}\frac{z_0}{\sqrt{1-z_0^2}} \ .
\label{21}
\end{eqnarray}
Similarly the deficit angle for $\phi_1$ is
\begin{equation}
\Delta \phi_1 = 2~ \frac{a\omega_2 q + b}{\Omega_2}~
\int_{\theta_0}^{\frac{\pi}{2}}~d\theta~\frac{\cot\theta}{\sqrt{\sin^2\theta
- \sin^2\theta_0}} =  \pi-\frac{\theta_0}{2}. \label{22}
\end{equation}
Now using equations (\ref{17}),(\ref{20}),(\ref{21}) and
(\ref{22}), we can rewrite the equation (\ref{14}) as magnon
dispersion relation
\begin{eqnarray}
(E - J_1)_{reg} =  &-& \sqrt{{(S-qT\Delta t)}^2_{reg} +
4T^2(1-q^2)\cos^2{\frac{\Delta t}{2}}} \cr & \cr && +
\sqrt{(J_2-qT\Delta\phi_1)^2 + 4T^2(1-q^2)\sin^2{\frac{\Delta
\phi_1}{2}}}. \label{23}
\end{eqnarray}
Note that by putting $q=0$ we get back the dispersion relation for
the three spin giant magnon solution presented in
\cite{Ryang:2006yq}. The above dispersion relation seemingly gives
us a collection of magnon bound states with momentum
$\Delta\phi_1$ with another collection of magnon bound states with
momentum $\Delta t + \pi$. However here both the number of magnons
in the bound states are somehow shifted by the `kink' charges
proportional to the `momenta'. The total dispersion relation is
completely non-periodic in these momenta owing to the presence of
the terms as explained above. For the WZW limit it can be expected
that naively putting $q = 1$ into the dispersion relation might
not be useful. The fate of these open string solutions in that
limit has to be investigated in a more rigorous way.

\subsection{Spike Solution}

To obtain the single spike-like solution, we chose the integration
constants as: $A_1 = \frac{a^2}{b} = A_3$  and $A_2 = 0 = A_4$.
The solution of equations (\ref{6}) and (\ref{7}) now becomes
\begin{equation}
\rho_y = \frac{\Omega_1}{a^2-b^2}\tanh\rho \sqrt{\cosh^2\rho -
\cosh^2\rho_1} \ , \label{24}
\end{equation}
and
\begin{equation}
\theta_y = \frac{\Omega_2}{a^2-b^2}\cot\theta \sqrt{\sin^2\theta -
\sin^2\theta_1} \ , \label{25}
\end{equation}
where, $\cosh\rho_1 = \frac{a^2-ab\omega_1 q}{b\Omega_1}$, and
$\sin\theta_1 = \frac{a^2+ab\omega_2 q}{b\Omega_2}$. Now the
conserved charges (\ref{8}) and deficit angles become
\begin{eqnarray}
E &=& \frac{T}{a^2-b^2}\int d\sigma~ a^2(1-q^2)\sinh^2\rho, \cr &
\cr \frac{S}{\omega_1} &=& \frac{T}{a^2-b^2}\int d\sigma~
\left[a^2(1-q^2) +
(a^2q^2-\frac{a^3q}{b\omega_1})\frac{1}{\cosh^2\rho}\right]\sinh^2\rho,
\cr & \cr J_1 &=& -\frac{T}{a^2-b^2}\int d\sigma~
a^2(1-q^2)\cos^2\theta , \cr & \cr \frac{J_2}{\omega_2} &=&
\frac{T}{a^2-b^2} \int d\sigma~ \left[a^2(1
-q^2)+(a^2q^2+\frac{a^3q}{b\omega_2})
\frac{1}{\sin^2\theta}\right]\cos^2\theta, \cr & \cr \Delta\phi_1
&=& \int dy ~~\left[\frac{a^2+ab\omega_2 q}{a^2b-b^3}
~\frac{\cos^2\theta}{\sin^2\theta} + \frac{1}{b}\right], \cr & \cr
\Delta t &=& \int dy ~~\left[\frac{ab\omega_1 q-a^2}{a^2b-b^3} ~
\frac{\sinh^2\rho}{\cosh^2\rho} + \frac{1}{b}\right]. \label{26}
\end{eqnarray}
From (\ref{26}), we get the follwing spike-like relation between
the conserved charges
\begin{equation}
\omega_1 E - S - q T(\Delta\phi_1 -\Delta t) =
2qT(\frac{\pi}{2}-\theta_1).  \label{27}
\end{equation}
For the sake of completeness we wish to compute $J_1$ and $J_2$ as
\begin{eqnarray}
J_1 &=& 2T~ \sqrt{\frac{1-q^2}{1-\omega^2_2}} ~\cos\theta_1, \cr &
\cr J_2 &=& -2T\omega_2~ \sqrt{\frac{1-q^2}{1-\omega^2_2}}
~\cos\theta_1 - q T (\pi-2\theta_1). \label{28}
\end{eqnarray}
From (\ref{28}), we have the follwing relation between $J_1$ and
$J_2$
\begin{equation}
\omega_2 J_1 + J_2= qT(2\theta_1-\pi). \label{29}
\end{equation}
\subsection{Circular Strings}
In this section, we describe the rotating circular string which is
rotating at a fixed $\rho$ value. For this we choose the
integration constants as $A_4=0$ and $A_3=b$. Now form the
equation (\ref{5}) we can get $A_1=A_2 \omega^2_1+b$. Using above
constant values in the equations (\ref{4}) and (\ref{7}), we get
\begin{equation}
\theta_y =
\frac{\cot\theta}{a^2-b^2}\sqrt{a^2(1-q^2)(1-\omega^2_2)}\sqrt{\sin^2\theta-\alpha^2},
\label{30}
\end{equation}
where $\alpha = \frac{b+aq\omega_2}
{\sqrt{a^2(1-q^2)(1-\omega^2_2}}$ and $\theta$ value runs from
$\frac{\pi}{2}$ to $\theta_{max} = \sin^{-1}\alpha$. The solution
for the above equation is
\begin{equation}
\cos\theta = \frac{\sqrt{1-\alpha^2}}{\cosh\beta y}, \label{31}
\end{equation}
where $\beta =
\frac{1}{a^2-b^2}\sqrt{a^2(1-q^2)(1-\omega^2_2)-(b+aq\omega_2)^2}$.
From the solution, we can see that at $\tau=0$, $\theta =
\frac{\pi}{2}$ corresponds to $\sigma=\pm\infty$ and
$\theta_{max}$ corresponds to $\sigma = 0$, which shows the range
of sigma to be $-\infty < \sigma < \infty$. Now the solution to
equation (\ref{6}) can be written as
\begin{equation}
(a^2-b^2)\rho^2_y = a^2(1-q^2)(1-\omega^2_1)\sinh^2\rho +
\frac{(A_1-aq\omega_1)^2} {\cosh^2\rho}-\frac{A^2_2 \omega^2_1}
{\sinh^2\rho} + C, \label{32}
\end{equation}
where $C$ is the integration constant can be chosen from the
$T_{\tau\sigma} = 0$ Virasoro constraint which is
\begin{equation}
C = 2aq\omega_1(A_1-A_2) - a^2q^2\omega^2_1 - b^2. \label{322}
\end{equation}
Now, we can write the equation(\ref{32}) as
\begin{equation}
\rho_y = \pm \frac{A}{(a^2-b^2)\cosh\rho \sinh\rho},\label{33}
\end{equation}
where
\begin{eqnarray}
A &=& \sqrt{M \sinh^6\rho + N \sinh^4\rho + P \sinh^2\rho + R},
\cr && \cr M &=& a^2(1-q^2)(1-\omega^2_1), \cr && \cr N &=&
a^2(1-q^2-\omega^2_1)-b^2+ 2aq\omega_1[b+A_2(\omega^2_1-1)], \cr
&& \cr P &=& 2bA_2 \omega^2_1 - 2aq \omega_1 A_2 + \omega^2_1
A^2_2(\omega^2_1 - 1), \cr && \cr R &=& -A^2_2 \omega^2_1.
\label{34}
\end{eqnarray}
The above equation can be written as
\begin{equation}
A = \sqrt{(x-r_1)(x-r_2)(x-r_3)}, \label{35}
\end{equation}
where $x=\sinh^2\rho$ and $r_{i=1,2,3} = r_i(a,b,q,\omega_1,A_2)$
are the roots of the above polynomial. The equation (\ref{35})
gives the minimum value for $\rho_{min} = \sinh^{-1}\sqrt{r_1}$,
where $r_1$ is the smallest root of the polynomial which
corresponds to the minimum value of $AdS$ radius i.e.
$\rho_{min}$. Using equation (\ref{4}) and (\ref{33}), we can
write
\begin{equation}
\frac{\partial \rho}{\partial \phi} = \frac{A \sinh\rho}{[A_2
\omega_1 + (aq - b \omega_1)\sinh^2\rho]\cosh\rho}. \label{36}
\end{equation}
The above differential equation describes the shape of the string
on $AdS$ part. The simple solution of the above equation (\ref{36})
is given by the string located at $\rho=\rho_{min}$ where $A$ is
zero. From the equation(\ref{4}) at a fixed time $\tau=0$ where
$y=a\sigma$, the string configuration in $\phi$-direction is given
by
\begin{equation}
\phi =
\frac{a\omega_1}{a^2-b^2}\left[\frac{A_2}{r_1}+\frac{aq}{\omega_1}-b\right]\sigma.
\label{37}
\end{equation}
We can notice that $\phi$ ranges as $\sigma$, which is $-\infty <
\phi < \infty$. So the solution corresponds to the circular string
having infinite number of windings. The conserved charges and deficit
angles for this configuration are given by
\begin{eqnarray}
E &=& \frac{T}{a^2-b^2} \int~
d\sigma~\left[a^2(1-q^2)\sinh^2\rho_{min} + a^2 - b^2 -A_2 b
\omega^2_1 - A_2aq\omega_1\right], \cr && \cr \frac{S}{\omega_1}
&=& \frac{T}{a^2-b^2} \int~
d\sigma~\left[a^2(1-q^2)\sinh^2\rho_{min} -bA_2-(A_2aq\omega_1 +
\frac{abq} {\omega_1}-a^2q^2)\frac{\sinh^2\rho_{min}}
{\cosh^2\rho_{min}}\right], \cr && \cr J_1 &=& \frac{T}{a^2-b^2}
\int~ d\sigma~[a^2-b^2-a^2(1-q^2)\cos^2\theta], \cr && \cr
\frac{J_2}{\omega_2} &=& \frac{T}{a^2-b^2} \int~ d\sigma~
\left[a^2(1-q^2)\cos^2\theta +
(a^2q^2+\frac{abq}{\omega_2})\frac{\cos^2\theta}{\sin^2\theta}
\right], \cr && \cr \Delta \phi_1 &=& \int ~ d\sigma~
\frac{b+aq\omega_2} {a^2-b^2} a \cot^2\theta, \cr && \cr \Delta t
&=& \int ~ d\sigma~ \frac{a}{a^2-b^2}\left[\frac{A_2
\omega^2_1}{\cosh^2\rho_{min}}-(b - aq
\omega_1)\frac{\sinh^2\rho_{min}} {\cosh^2\rho_{min}}\right].
\label{38}
\end{eqnarray}
Now choosing $\omega^2_1 = 1-b^2$ and $A_2 = \frac{2}{b}$, from
the above equations we can see the relation among the conserved
quantities as
\begin{equation}
E - \frac{S-qT\Delta t}{\omega_1} =
\frac{a^2+b^2}{a^2-b^2}\left[J_1+\frac{J_2-qT\Delta
\phi_1}{\omega_2}\right]. \label{39}
\end{equation}
Note that the above relation reduces to the one mentioned in
\cite{Lee:2008sk} in the limit $q=0$.
\subsection{Infinite spikes on AdS}
The equation (\ref{36}) also describes another kind of string
solution where the string extends from $\rho_{min}$ to $\infty$
with the infinite winding number and the infinite angular momentum
in the $\phi$ direction. As the string is now extended in the
radial direction of $AdS$, $\rho$ is not fixed anymore and becomes
the function of $\sigma$ (as we are considering the case at $\tau
= 0$). In the asymptotic region, the solution to the equation
(\ref{33}) is in the form $\sigma - C \sim e^{-\rho}$, where $C$
is an integration constant. So when $\rho \rightarrow \infty$,
$\sigma$ goes to $C$, which tells us $\rho_{min} < \rho < \infty$
covers the finite range of $\sigma$ which corresponds to that of
one AdS spike solution. To cover all $\sigma$, we must include an
array of infinite spikes on the AdS. As the integral range of
$\rho$ covers only one spike, we can write conserved charges and
deficit angles for a single AdS spike as
\begin{eqnarray}
E &=& 2 T \int_{\rho_{min}}^{\infty}
d\rho\left[a(1-q^2)\sinh^2\rho + a-\frac{b^2}{a} -
A_2b\frac{\omega_1}{a} -A_2q\omega_1\right] \frac{\cosh\rho
\sinh\rho}{A}, \cr && \cr \frac{S}{\omega_1} &=& 2T
\int_{\rho_{min}}^{\infty} d\rho\left[a(1-q^2)\sinh^2\rho
-\frac{b}{a}A_2 -(A_2q\omega_1+
\frac{bq}{\omega_1}-aq^2)\tanh^2\rho \right] \frac{\cosh\rho
\sinh\rho}{A}, \cr && \cr \Delta t &=& 2
\int_{\rho_{min}}^{\infty} d\rho\left[\frac{A_2
\omega^2_1}{\cosh^2\rho} - (b-aq\omega_1) \tanh^2\rho \right]
\frac{\cosh\rho \sinh\rho}{A}, \cr && \cr \Delta \phi &=& 2
\int_{\rho_{min}}^{\infty} d\rho\left[ A_2 \omega_1 +(aq
-b\omega_1)\sinh^2\rho \right]\frac{\coth\rho}{A}. \label{40}
\end{eqnarray}
We use the same value of $A_2= 2/b$ and $\omega^2_1= 1-b^2$. From
the above conserved charges we can find the relation as
\begin{equation}
E - \frac{S-qT\Delta t}{\omega_1} = 2T
\frac{a^2+b^2}{a^2q-ab\omega_1}\left[\frac{\Delta
\phi}{2}-I\right], \label{41}
\end{equation}
where
\begin{eqnarray}
I &=& A_2 \omega_1\int_{\rho_{min}}^{\infty} d\rho
\frac{\coth\rho}{A} = \frac{\omega_1}{b} \int_{r_1}^{\infty} dx
\frac{1}{x \sqrt{(x-r_1)(x-r_2)(x-r_3)}} \cr && \cr &=&
\frac{2\omega_1}{b r_3 \sqrt{r_3-r_1}}\left[\Pi
\left(\frac{r_3}{r_3-r_1}, \frac{r_3-r_2}{r_3-r_1}\right) -
\mathbb{K}\left(\frac{r_3-r_2}{r_3-r_1}\right)\right], \label{42}
\end{eqnarray}
where $\Pi$ and $\mathbb{K}$ are complete elliptical integrals of
third and first kind and \\ $\Pi (n,k) =
\int_{0}^{1}\frac{dx}{(1+nx^2)\sqrt{(1-x^2)(1-k^2x^2)}}$. Now as
$\theta$ covers all the ranges of $\sigma$, we can write the
conserved charges and deficit angles with the help of $\theta$
equation where $\rho$ is a complicated function of $\theta$,
\begin{equation}
\frac{d \rho}{d \theta} = \frac{A
\tan\theta}{\sqrt{a^2(1-q^2)(1-\omega^2_2)} \cosh\rho \sinh\rho
\sqrt{\sin^2\theta - \alpha^2}}. \label{43}
\end{equation}
Now the conserved charges and deficit angles can be written by the
help of the above equation (\ref{43})
\begin{eqnarray}
E &=& \frac{2T}{a^2\Lambda} \int_{\pi/2}^{\theta_{max}} d\theta
\left[a^2(1-q^2)\sinh^2\rho + a^2 - b^2 -A_2 b \omega^2_1 -
A_2aq\omega_1\right]\frac{\tan\theta}{\sqrt{\sin^2\theta-\alpha^2}},
\cr && \cr \frac{S}{\omega_1} &=& \frac{2T}{a^2\Lambda}
\int_{\pi/2}^{\theta_{max}} d\theta \left[a^2(1-q^2)\sinh^2\rho
-bA_2-(A_2aq\omega_1 + \frac{abq}
{\omega_1}-a^2q^2)\frac{\sinh^2\rho} {\cosh^2\rho}\right]
\frac{\tan\theta}{\sqrt{\sin^2\theta-\alpha^2}}, \cr && \cr J_1
&=& \frac{2T}{a^2\Lambda} \int_{\pi/2}^{\theta_{max}} d\theta
[a^2-b^2-a^2(1-q^2)\cos^2\theta]
\frac{\tan\theta}{\sqrt{\sin^2\theta-\alpha^2}}, \cr && \cr
\frac{J_2}{\omega_2} &=& \frac{2T}{a^2\Lambda}
\int_{\pi/2}^{\theta_{max}}  d\theta \left[a^2(1-q^2)\cos^2\theta
+ (a^2q^2+\frac{abq}{\omega_2})\frac{\cos^2\theta}{\sin^2\theta}
\right] \frac{\tan\theta}{\sqrt{\sin^2\theta-\alpha^2}}, \cr &&
\cr \Delta \phi_1 &=& 2 \frac{b+aq\omega_2} {a\Lambda}
\int_{\pi/2}^{\theta_{max}} d\theta \cot^2\theta
\frac{\tan\theta}{\sqrt{\sin^2\theta-\alpha^2}}, \cr && \cr \Delta
t &=& \frac{2}{a\Lambda} \int_{\pi/2}^{\theta_{max}}  d\theta
\left[\frac{A_2 \omega^2_1}{\cosh^2\rho_{min}}-(b - aq
\omega_1)\frac{\sinh^2\rho}
{\cosh^2\rho}\right]\frac{\tan\theta}{\sqrt{\sin^2\theta-\alpha^2}},
\label{44}
\end{eqnarray}
where $\Lambda = \sqrt{(1-q^2)(1-\omega^2_2)}$. These charges for
the array of infinite spikes also satisfy the same relation as in
the equation (\ref{39}), which can be rewritten as
\begin{equation}
E - S^{\prime} - {J^{\prime}}_1 = 2T \frac{a^2+b^2}{a^2-b^2}
\sqrt{\frac{1-q^2}{1-\omega^2_2}} \sin{\frac{\Delta \phi_1}{2}},
\label{45}
\end{equation}
where
\begin{eqnarray}
S^{\prime} &=& \frac{S-qT\Delta t}{\omega_1},~~~ {J^{\prime}}_1 =
\frac{a^2+b^2}{a^2-b^2} J_1,~~~ \Delta \phi_1 = 2
\theta_{max} - \pi \nonumber \\
&&\frac{J_2-qT\Delta \phi_1}{\omega_2} = - 2T \
\sqrt{\frac{1-q^2}{1-\omega^2_2}} \cos\theta_{max} .\nonumber
\end{eqnarray}
This also looks analogous to the relation mentioned in
\cite{Lee:2008sk} with $q = 0$. In the limit of $b\rightarrow 0$
and $\sin{(\Delta \phi_1 /2)} \rightarrow 1$, the above relation
(\ref{45}) becomes
\begin{equation}
E - S^{\prime} - J_1 = 2T \sqrt{\frac{1-q^2}{1-\omega^2_2}}.
\label{46}
\end{equation}
The above relation (\ref{46}) looks like circular string rotating
at $\rho_{min}$ with the infinite angular momentum $S$ which is
shifted by the kink charge and has the shape of a magnon on $S^2$.
Interestingly we also get the same relation (\ref{46}) from the
circular string equation (\ref{39}) using the same limit of $b$
and $\Delta \phi_1$. So both the circular string and infinite
spikes resemble the shape of a magnon solution in the $AdS_3
\times S^2$ , when we use the specific limits.
\section{Conclusions and Outlook}
In this paper, we have analyzed a large variety of semiclassical
string solutions in $AdS_3\times S^3$ motivated mainly by recent
studies on string theory in $AdS_3\times S^3\times T^4$ with mixed
fluxes. First we have studied the strings in $\mathbb{R}\times
S^3$ with a background NS-NS flux. In \cite{Hoare:2013ida} the
generalization of the known dyonic giant magnon solution was found
using a WZ term describing the NS-NS flux in a principal chiral
model. The `shift term' in the angular momentum proportional to
magnon momenta $p$ was described to be arising due to the
ambiguity in the conserved charges associated to the boundary
terms of the theory. In our case we find that out from classical
string solutions, since it is equivalent to the chiral model in
the conformal gauge. We also found out the single spike solution
is modified due the background field. While the spike relation
presented in \cite{Ishizeki:2007we} does not change, the
dispersion relation between the angular momenta $J_1$ and $J_2$
appear to be different. In the last section, we have investigated
various classes of rotating open string solutions for the string
moving on $AdS_3\times S^3$ with one spin in the $AdS$ space and
two angular momenta on the sphere. Here the NS-NS fluxes
parameterized by $q$ are turned on in both $AdS$ and the sphere.
The three spin giant magnon solution in this case is also found to
be modified by the `shift terms' already presented in the previous
case. While the angular momenta part is modified by a term
proportional to the dyonic giant magnon momenta as before, the
$AdS$ spin part seems to modified by a kink charge proportional to
$\Delta t$. We also discuss a spike-like solution in this
background. Further following \cite{Lee:2008sk} we found the
circular string solutions and an array of infinitely many spikes
in the $AdS$. We presented the generalized dispersion relations
between conserved quantities in these cases for some specific
choice of constants and showed that they reduce to the relations
presented in \cite{Lee:2008sk} when the B-fields are turned off.
It would be challenging to identify the dual gauge theory
operators corresponding to these string solutions presented in
section-2, though we can not expect a spin chain interpretation of
the integrable system for $q \neq 0$ due to the non-periodicity of
the solution in the momentum $p$. It is to be noted that all our
solutions reduce to the ones known in the literature for the $q =
0$ case. However $q=1$ appears to be a special case as world sheet
theory is related to a WZW model, and it will be interesting to
investigate the behaviour of our solutions in this limit. Also
since $AdS_3\times S^3$ with NS-NS flux can be represented as a
$SL(2,R)\times SU(2)$ WZW model, we can try to find the three spin
giant magnon solution following the formulation in
\cite{Hoare:2013ida} and comment on the ambiguity in the $B$ field
for the $AdS$ part also. The three spin string solutions presented
in section 3 are related to the open strings and are interesting
by themselves. However, apriori it is not to us clear whether one
can learn about them in the boundary theory by using AdS/CFT
duality. Furthermore it will be interesting to investigate the
finite size effects on the solutions presented here as it is shown
in \cite{Ahn:2014tua} that the leading finite size effects in
$\mathbb{R}\times S^3$ with a $B$-field is quite different from
the usual ones. We wish to come back to these issues in near
future.

\section*{Acknowledgement} We would like to thank A. Tseytlin for a
correspondence. KLP would like to thank CERN PH-TH for generous
hospitality and financial support where a part of this work was
done.

\end{document}